\documentclass{article}
\usepackage[utf8]{inputenc}

\usepackage{graphicx}
\usepackage{authblk}
\usepackage[utf8]{inputenc}
\usepackage{mathptmx}
\usepackage{epsfig,amsopn}
\usepackage{graphicx}
\usepackage{braket}
\usepackage{float}
\usepackage{enumerate}
\usepackage[normalem]{ulem}
\usepackage{todonotes}

\begin{document}

\title{Mean-performance of sharp restart I:\\ Statistical roadmap}
\author[1]{Iddo Eliazar \thanks{email: eliazar@tauex.tau.ac.il}}
\author[1]{Shlomi Reuveni \thanks{email: shlomire@tauex.tau.ac.il}}
\affil[1]{School of Chemistry, Center for the Physics and Chemistry of Living Systems, The Sackler Center for Computational Molecular and Materials Science, and The Mark Ratner Institute for Single Molecule Chemistry, Tel Aviv University, Tel Aviv 6997801, Israel}
\maketitle

\begin{abstract}
\

Restart is a general framework, of prime importance and wide applicability, for expediting first-passage times and completion times of general stochastic processes. Restart protocols can use either deterministic or stochastic timers. Restart protocols with deterministic timers -- ``sharp restart'' -- assume a principal role: if there exists a restart protocol that improves mean-performance, then there exists a sharp-restart protocol that performs as good or better. This paper, the first of a duo, presents a comprehensive mean-performance analysis of sharp restart. Using statistical methods, the analysis establishes universal criteria that determine when sharp restart improves or worsens mean-performance, i.e., decreases or increases mean first-passage/completion times. These criteria are akin to those recently discovered for the most widely applied restart protocols -- ``exponential restart'' -- which use exponentially-distributed timers. However, while the exponential-restart criteria cover only the case of slow timers, the sharp-restart criteria established here further cover the cases of fast, critical, and general timers; moreover, the latter criteria address the very existence of timers with which sharp restart improves or worsens mean-performance. Using the slow-timers criteria, we discover a general scenario for which: sharp restart improves mean-performance, whereas exponential restart worsens mean-performance. The potency of the novel results presented here is demonstrated by examples, and by the results' application to canonical diffusion processes.\\


\end{abstract}

\newpage

\section{\label{1}Introduction}

Restart -- also known as ``resetting'' -- of stochastic processes is a subject that drew vigorous scientific investigation recently \cite{review}-\cite{Open12}. Of particular interest in that regard is an ongoing effort to characterize and understand the effect restart has on first-passage times and on completion times of stochastic processes \cite{FPUR0}-\cite{FPUR9}. For example, think of a randomized computer algorithm that failed to converge within a given time frame. Should the algorithm be allowed to continue to run, or would it be better to halt and start a new run of the algorithm \cite{CS1}-\cite{CS4}? A similar dilemma faces foraging animals that are searching for food, and, more generally, agents that are searching for a target. If the search is unsuccessful for a while -- should it be continued, or is it better to stop and start the search afresh? Taking an informed decision is critical, as it may very well spell the difference between a winning and losing search strategy \cite{HRS,pollination}.

Restart is also important at the molecular scale, where it is an integral part of a variety of chemical reactions and biological processes \cite{Restart-Mol1}-\cite{Restart-Mol8}. For example, in enzymatic catalysis the formation of an enzyme-substrate complex is a necessary step en-route to product generation. Unbinding of an enzyme from a substrate resets the enzymatic reaction. In some circumstances this resetting impedes product generation, whereas in other circumstances the resetting expedites product generation \cite{FPUR0,Restart-Mol1,Restart-Mol3,Restart-Mol6}. Similarly, molecular-biology search processes -- e.g., transcription factors seeking their DNA target sites -- can be either slowed-down or sped-up by unbinding \cite{Restart-Mol_Search1,Restart-Mol_Search2}. In these examples, and with regard to general running processes, it is crucial to understand the effect of restart: when is it impeding, and when is it expediting?

A given stochastic process can be restarted by any one of infinitely many restart protocols \cite{FPUR1}-\cite{FPUR3},\cite{FPUR8,HRS},\cite{General_restart1}-\cite{General_restart6}. Arguably, the most common and widely applied restart protocol is ``exponential restart'' -- which uses stochastic timers that are exponentially distributed. Specifically, in exponential restart the resetting epochs follow Poisson-process statistics. The effect of exponential restart on the first-passage times of stochastic processes was extensively studied for, e.g.: diffusions \cite{freeD1}-\cite{ScaledBM}; diffusions in confined domains \cite{ConD1,ConD2}; diffusions in various potentials \cite{PotD1}-\cite{PotD3}; motions with L\'evy flights \cite{Levy1,Levy2}; continuous time random walks \cite{CTRW}; and telegraph processes \cite{Telegraph}. Going beyond particular examples, a universal criteria for exponential restart with slow timers (i.e. exponentially-distributed timers with large means) was recently established \cite{FPUR0,FPUR1,Restart-Mol1,Optimal_fluc}: based on the coefficient of variation (CV) of a first-passage time of a given stochastic process, the criteria determine when restart increases or decreases the mean first-passage time.

Consider now restart protocols that use general stochastic timers with a given positive mean. Via their timers, entropy quantifies the protocols' inherent randomness. Maximal entropy is attained by exponentially distributed timers -- and hence exponential restart assumes the role of the `most random' restart protocol. Conversely, minimal entropy is attained by deterministic timers -- and hence ``sharp restart'' assumes the role of the `least random' restart protocol. Sharp restart is of principal importance due to the following key result \cite{FPUR1,FPUR2}: with regard to mean performance, sharp restart either matches or outperforms any other restart protocol. Namely, if there is a restart protocol that improves the mean first-passage/completion time (of a given stochastic process), then there exists a sharp-restart protocol that attains at least as good an improvement.

Counter-wise to its principal importance, the investigation of sharp restart in the physics literature is rather limited. A few particular examples of sharp restart were worked out \cite{Sharp1,Sharp2}. However, universal criteria for sharp restart -- akin to the aforementioned CV criteria for exponential restart -- are not available. This paper is the first part of a duo addressing the mean-performance of sharp restart in detail: here we present a comprehensive statistical analysis, and in the second part we shall present a comprehensive analysis based on a socioeconomic-inequality perspective. Specifically, this paper establishes six universal criteria that determine when sharp restart improves or worsens mean first-passage/completion times. The novel criteria regard sharp restart with various timers: fast, slow, critical, and general. Also, the novel criteria address the very existence of timers with which sharp restart improves or worsens mean first-passage/completion times.

Comparing exponential restart to sharp restart, we point out that: while the universal exponential-restart criteria cover only the case slow timers, the universal sharp-restart criteria presented here cover many more types of timers. Moreover, regarding slow timers, in this paper we discover a general scenario in which: exponential restart worsens mean first-passage/completion times, whereas sharp restart improves them. This general scenario, as well as other key results established in this paper, are illustrated by examples. In particular, the examples demonstrate the results' application to canonical diffusion processes.

The remainder of the paper is organized as follows. Section \ref{2} provides a concise description of sharp restart. Section \ref{3} uses the renewal-processes notion of residual lifetime to determine, for any given sharp-restart timer, if mean-performance improves or worsens. Section \ref{8} \ further uses the notion of residual lifetime to determine the very existence of sharp-restart timers for which mean-performance improves or worsens. Section \ref{4} uses the reliability-engineering notion of hazard rate to investigate the global optimum -- as a function of the sharp-restart timer -- of mean-performance. As shall be shown, the global optimum can be attained either by infinitely fast timers, or by infinitely slow timers, or by critical timers. Section \ref{5} further explores the case of fast and slow timers, and Section \ref{6} further explores the case of critical timers. Section \ref{7} concludes with a summary of the six universal sharp-restart criteria, and with a discussion: comparison to the universal exponential-restart criteria; the aforementioned general scenario; applications to canonical diffusion processes; and a brief outlook.

A note about notation: $\mathbf{E}\left[ X \right] $ denotes the expectation of a (non-negative) random variable $X $; and IID is acronym for independent and identically distributed (random variables).

\section{\label{2}Sharp restart}

\emph{Sharp restart} is an \emph{algorithm} that is described as follows. There is a general task with completion time $T$, a positive-valued random variable. To this task a three-steps algorithm, with a positive deterministic timer $\tau $, is applied. Step I: initiate simultaneously the task and the timer. Step II: if the task is accomplished up to the timer's expiration -- i.e. if $T\leq \tau $ -- then stop upon completion. Step III: if the task is not accomplished up to the timer's expiration -- i.e. if $T>\tau$ -- then, as the timer expires, go back to Step I.

The sharp-restart algorithm generates an iterative process of independent and statistically identical task-completion trials. This process halts during its first successful trial; following convention, we denote by $T_{R}$ the halting time of the process \cite{FPUR1}. Namely, $T_{R}$ is the overall time it takes -- when the sharp-restart algorithm is applied -- to complete the task. The sharp-restart algorithm is a \emph{non-linear mapping} whose input is the random variable $T$, whose output is the random variable $T_{R}$, and whose parameter is the deterministic timer $\tau$.

The input-to-output map $T\mapsto T_{R}$ admits the stochastic representation
\begin{equation}
T_{R}=I \left\{ T\leq \tau \right\} \cdot T+I\left\{ T>\tau \right\} \cdot \left( \tau +T_{R}^{\prime }\right) ,  \label{21}
\end{equation}
where: $I\left\{ E\right\} $ is the indicator function of the event $E$;\footnote{Namely: $I\left\{ E\right\} =1$ if the event $E$ occurred, and $I\left\{ E\right\} =0$ if the event $E$ did not occur.} and $T_{R}^{\prime }$ is an IID copy of the output $T_{R}$ that is independent of the input $T$. Indeed, if the event $\left\{ T\leq \tau \right\} $ occurs then $T_{R}=T$. And, if the event $\left\{ T>\tau \right\} $ occurs then $T_{R}=\tau +T_{R}^{\prime }$ i.e.: $\tau $ time units are spent on the first (unsuccessful) task-completion trial; the task-completion process starts anew at the end of the first (unsuccessful) trial; and then an additional $T_{R}^{\prime }$ time units are spent till the new task-completion process halts.

In this paper we use the following notation regarding the input's statistics: distribution function, $F\left( t\right) =\Pr \left( T\leq t\right) $ ($t\geq 0$); survival function, $\bar{F}\left( t\right) =\Pr \left( T>t\right) $ ($t\geq 0$); density function, $f\left( t\right) =F^{\prime }\left( t\right) =-\bar{F}^{\prime }\left( t\right) $ ($t>0$); and mean, $\mu =\mathbf{E}\left[ T\right] =\int_{0}^{\infty }tf\left(t\right) dt$. With no loss of generality, the input's density function is henceforth assumed to be positive-valued over the positive half-line: $f\left( t\right) >0$ for all $t>0$.\footnote{This is merely a technical assumption, which is introduced in order to assure that all positive timers $\tau$ are admissible. In general, admissible timers are in the range $t_{low} < \tau < \infty$, where $t_{low}$ is the lower bound of the support of the input's density function: $t_{low}=\inf \left\{ t>0 \:|\: f\left( t\right) >0\right\}$.}

We denote by $M\left( \tau \right) =\mathbf{E}\left[ T_{R}\right] $ the output's mean; this notation underscores the fact that the output's mean is a function of the timer $\tau $, the parameter of the sharp-restart algorithm. In terms of the input's distribution and survival functions, the output's mean is given by \cite{FPUR1,BP2}:

\begin{equation}
M\left( \tau \right) =\frac{1}{F\left( \tau \right) }\int_{0}^{\tau }\bar{F} \left( t\right) dt.  \label{22}
\end{equation}%
Eq. (\ref{22}) is obtained by taking expectation on both sides of Eq. (\ref {21}), and thereafter performing a probabilistic calculation. The derivation of Eq. (\ref{22}) is detailed in the Methods.

A key issue in the context of the sharp-restart algorithm is determining when will its application expedite task-completion, and when will it impede task-completion. To address this issue, the main approach employed in the physics literature is \emph{mean-performance}: comparing the input's mean $\mu $ to the output's mean $M\left( \tau \right) $, and checking which of the two means is smaller \cite{FPUR0}-\cite{FPUR3},\cite{FPUR5}-\cite{HRS}. To that end we use the following
terminology:

\begin{enumerate}
\item[$\bullet $] Sharp restart with timer $\tau $ is \emph{beneficial} if it improves mean-performance, $M\left( \tau \right) <\mu $.

\item[$\bullet $] Sharp restart with timer $\tau $ is \emph{detrimental} if it worsens mean-performance, $M\left( \tau \right) >\mu $.
\end{enumerate}

The mathematical-statistical setting that underpins restart is identical to the setting that underpins preventive maintenance \cite{BH}-\cite{Call}. However, while sharing the same setting, the two topics aim at different goals. As described above, in restart the goal is to expedite task-completion. On the other hand, in preventive maintenance the goal is to minimize long-term operating costs. There are some analogies between restart and preventive maintenance; yet, in general, the different goals of these topics lead to different analyses and results.

Eq. (\ref{22}) implies that the output's mean is always finite: $M\left(\tau \right) <\infty $ for all timers $\tau $. Thus, if the input's mean is infinite, $\mu =\infty $, then the application of the sharp-restart algorithm is highly beneficial -- as it reduces the input's infinite mean to the output's finite mean: $M\left( \tau \right) <\mu =\infty $. Having resolved the case of infinite-mean inputs, we henceforth set the focus on the case of finite-mean inputs, $\mu <\infty $. In the next sections we shall address the case of finite-mean inputs via two different perspectives: residual lifetimes and hazard rates.

\section{\label{3}Residual perspective}

Consider a renewal process \cite{Smi}-\cite{Ros} that repeats indefinitely, in a Sisyphean and independent manner, the underlying task whose completion time is the input $T$. This renewal process is described as follows.

We start at time $t=0$ and perform the task for the first time; upon the first completion we start performing the task for the second time; upon the second completion we start performing the task for the third time; and so on and so forth. Denoting by $\left\{ T_{1},T_{2},T_{3},\cdots \right\} $ the durations of the tasks -- these durations being IID copies of the input $T$ -- we obtain that: $T_{1}$ is the completion time of the first task, $T_{1}+T_{2}$ is the completion time of the second task, $T_{1}+T_{2}+T_{3}$ is the completion time of the third task, etc. The sequence of completion times is the renewal process generated from the input $T$.

Now, tracking the renewal process from some large time epoch \thinspace $t=l$ onwards, consider the waiting duration till observing the first task completion after time $t=l$. For a finite-mean input $T$, the theory of renewal processes asserts the following asymptotic result \cite{Ros}: the waiting duration converges in law, as $l\rightarrow \infty $, to a stochastic limit -- a positive-valued random variable $T_{res}$ that is termed the \emph{residual lifetime}\ of the input $T$. The density function governing the statistical distribution of the residual lifetime $T_{res}$ is \cite{Ros}:
\begin{equation} 
f_{res}\left( t\right) =\frac{1}{\mu}\bar{F}\left( t\right)  \label{31} 
\end{equation} 
($t\geq 0$). In turn, the distribution and survival functions of the residual lifetime $T_{res}$ are, respectively, $F_{res}\left( t\right)=\int_{0}^{t}f_{res}\left( u\right) du$ ($t\geq 0$) and $\bar{F}_{res}\left(t\right) =\int_{t}^{\infty }f_{res}\left( u\right) du$ ($t\geq 0$).

With the residual lifetime $T_{res}$ at hand, we can re-formulate Eq. (\ref{22}). Indeed, dividing both sides of Eq. (\ref{22}) by the input's mean $\mu $, and then using Eq. (\ref{31}) and the distribution and survival functions of the residual lifetime $T_{res}$, we arrive at the following formula for the ratio of the output's mean to the input's mean: 
\begin{equation}
\frac{M\left( \tau \right) }{\mu }=\frac{F_{res}\left( \tau \right) }{F\left( \tau \right) }=\frac{1-\bar{F}_{res}\left( \tau \right) }{1-\bar{F}\left( \tau \right) }.  \label{32}
\end{equation}
Note that the terms appearing on the middle and right part of Eq. (\ref{32}) depend only on the input's statistical distribution. Equation (\ref{32}) straightforwardly implies the following pair of \emph{residual criteria}:

\begin{enumerate}
\item[$\bullet $] Sharp restart with timer $\tau $ is beneficial if and only if $\bar{F}\left( \tau \right) <\bar{F}_{res}\left( \tau \right) $.

\item[$\bullet $] Sharp restart with timer $\tau $ is detrimental if and only if $\bar{F}\left( \tau \right) >\bar{F}_{res}\left( \tau \right) $.
\end{enumerate}

\begin{figure}[t!]
\centering
\includegraphics[width=12cm]{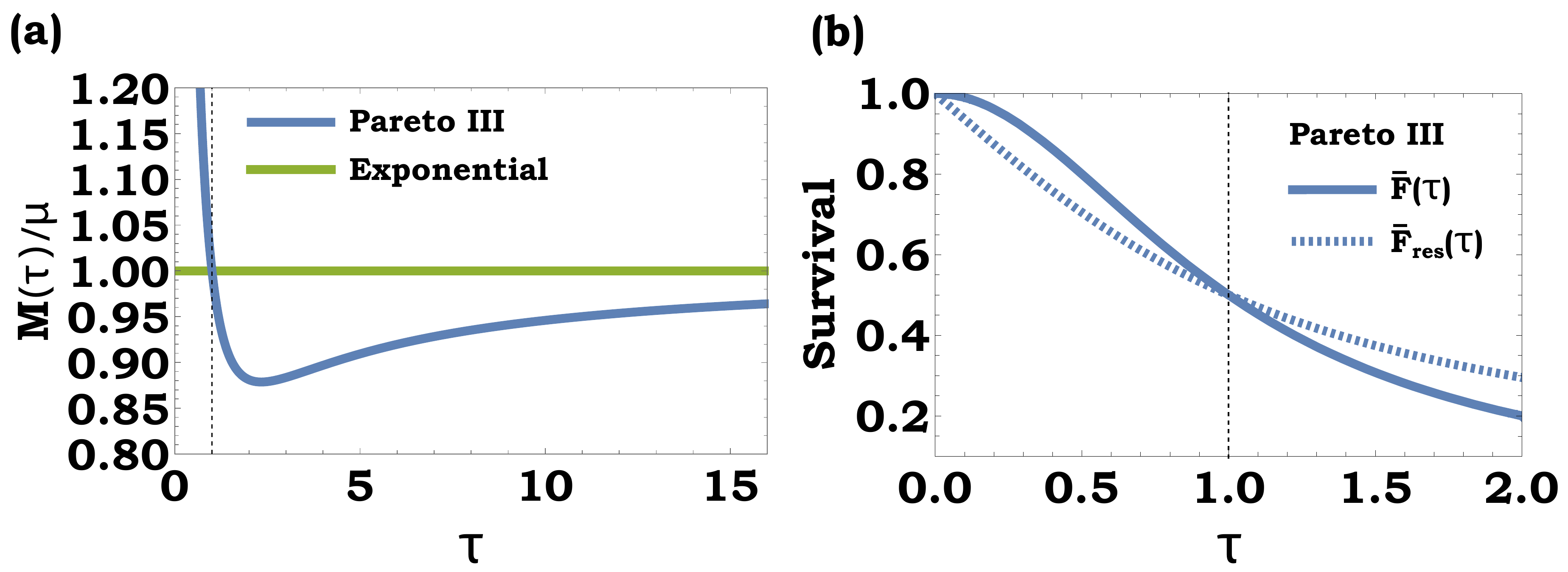}
\caption{Illustration of the pair of residual criteria. Panel (a): The ratio of the output’s mean $M(\tau)$ to the input's mean $\mu$ [Eq. (\ref{32})], plotted vs. the timer parameter $\tau$. Two plots are depicted: one for an Exponential input -- characterized by the survival function $\bar{F}\left( t\right)=\exp \left( -t/\mu \right)$; and one for a type-III Pareto input -- characterized by the survival function $\bar{F}\left( t\right)=1/(1+t^{p})$ (where the parameter $p$ is the Pareto power, and here $p=2$). For the Exponential input sharp restart is neither beneficial nor detrimental. For the Pareto type-III input sharp restart with timers $\tau>1$ is beneficial, and sharp restart with timers $\tau<1$ is detrimental. Panel (b): The survival functions $\bar{F}\left( \tau \right)$ and $\bar{F}_{res}\left( \tau \right)$ for the Pareto type-III input of panel (a). For timers $\tau>1$ we have $\bar{F}\left( \tau \right) <\bar{F}_{res}\left( \tau \right) $, and for timers $\tau<1$ we have $\bar{F}\left( \tau \right) >\bar{F}_{res}\left( \tau \right) $. Evidently, for the type-III Pareto input, panels (a) and (b) are in accord with the residual criteria.}
\label{Hazard_fig1}
\end{figure}

The pair of residual criteria determines the mean-performance of the sharp-restart algorithm by comparing the survival function of the input $T$ to the survival function of the input's residual lifetime $T_{res}$. We emphasize that the pair of residual criteria may provide different answers for different timers $\tau $. Indeed, it may be that while restart is beneficial for some timers $\tau $, it is detrimental for other timers $\tau$ and vice versa (Fig. 1).

A well-known fact from probability theory asserts that a finite-mean input $T$ is equal in law to its residual lifetime $T_{res}$ if and only if the input is Exponential \cite{Ros}. Specifically, an Exponential input $T$ is characterized by the exponential survival function $\bar{F}\left( t\right) =\exp \left( -t/\mu \right) $ ($t\geq 0$). In turn, combining this fact with Eq. (\ref{32}), we arrive at the following conclusion: $M\left( \tau \right) =\mu $ for all timers $\tau$ if and only if the input $T$ is Exponential (Fig. 1).

\section{\label{8}Existence results}

The  pair of residual criteria established in the previous section are timer-specific. Namely, for a given timer $\tau $, the pair of residual criteria determines if sharp restart with that specific timer is beneficial or detrimental. In this section we shift from timer-specific criteria to \emph{existence criteria}: results that determine the very existence of timers for which sharp restart is beneficial or detrimental. We note that existence criteria do not pinpoint specific timers $\tau $ for which sharp restart is beneficial or detrimental.

\subsection{\label{CV} Residual lifetime and CV}

The mean of a positive-valued random variable equals the integral, over the positive half-line, of its survival function. Hence, the mean of the input $T $ is given by $\mu =\int_{0}^{\infty}\bar{F}\left( t\right) dt$, and the mean of the input's residual lifetime $T_{res}$ is given by $\mu_{res}=\int_{0}^{\infty }\bar{F}_{res}\left( t\right) dt$. \ Also, in terms of the first and second moments of the input $T$, the mean of the input's residual lifetime $T_{res}$ is given by $\mu _{res}=\mathbf{E}[T^{2}]/(2\mu)$ \cite{Cox}-\cite{Ros}. Consequently, combining these facts together, a simple calculation yields 
\begin{equation}
\int_{0}^{\infty }\left[ \bar{F}\left( t\right) -\bar{F}_{res}\left(t\right) \right] dt=\mu -\mu _{res}=\frac{\mu ^{2}-\sigma ^{2}}{2\mu },\label{34}
\end{equation}%
where $\sigma $ is the input's standard deviation and $\sigma ^{2}$ is the
input's variance.

Comparing the input's mean $\mu $ to the mean $\mu _{res}$ of the input's residual lifetime -- or, equivalently, comparing the input's mean $\mu $ to the input's standard deviation $\sigma $ -- determines the existence of timers $\tau $ for which sharp restart is beneficial or detrimental. Indeed, Eq. (\ref{34}) together with the residual criteria of section \ref{3} yield the following pair of \emph{existence criteria}:

\begin{enumerate}
\item[$\bullet $] If $\mu <\mu _{res}$ -- which is equivalent to $\mu<\sigma $ -- then there exist timers $\tau $ for which sharp restart is
beneficial.

\item[$\bullet $] If $\mu >\mu _{res}$ -- which is equivalent to $\mu >\sigma $ -- then there exist timers $\tau $ for which sharp restart is detrimental.
\end{enumerate}

The existence criteria are similar to recently established criteria for exponential restart with slow timers \cite{FPUR0,FPUR1,HRS,Restart-Mol1,Optimal_fluc}. Specifically, with regard to exponentially-distributed timers with large means,\footnote{Or, described equivalently: exponential restart with small Poissonian resetting rates.} it was shown that: if $\sigma / \mu >1$ then exponential restart is beneficial; and if $\sigma / \mu <1$ then exponential restart is detrimental. As the ratio $\sigma / \mu$ is the input's coefficient of variation (CV), these exponential-restart criteria were referred to as ``CV criteria''.

\subsection{\label{Res} Residual lifetime and minimum}

Given two IID copies of the input, $T_{1}$ and $T_{2}$, denote by $f_{\max}\left( t\right) $ ($t>0$) the density function of the copies' maximum $\max\left\{ T_{1},T_{2}\right\} $, and denote by $\mu _{\min }$ the mean of the copies' minimum $\min \left\{ T_{1},T_{2}\right\} $. With these notations at hand, we can now present two equivalent formulae that follow from Eq. (\ref{32}); the derivations of these formulae are detailed in the Methods.

The first formula is
\begin{equation}
\int_{0}^{\infty }\left[ \frac{M\left( \tau \right) -\mu }{\mu }\right] f_{\max }\left( \tau \right) d\tau =2\Pr \left(T>T_{res}\right) -1,
\label{36}
\end{equation}%
where the random variables that appear in the right hand side of Eq. (\ref{36}) -- the input $T$ and the input's residual lifetime $T_{res}$ -- are independent of each other. The second formula is
\begin{equation}
\int_{0}^{\infty }\left[ M\left( \tau \right) -\mu \right] f_{\max }\left(\tau \right) d\tau =2\mu _{\min }-\mu .  
\label{37}
\end{equation}
Eq. (\ref{36}) and Eq. (\ref{37}) yield the following pair of \emph{existence criteria}:

\begin{enumerate}
\item[$\bullet $] If $\mu >2\mu _{\min }$ -- which is equivalent to $\Pr\left( T>T_{res}\right) <\frac{1}{2}$ -- then there exist timers $\tau $ for which sharp restart is beneficial.

\item[$\bullet $] If $\mu <2\mu _{\min }$ -- which is equivalent to $\Pr\left( T>T_{res}\right) >\frac{1}{2}$ -- then there exist timers $\tau $ for which sharp restart is detrimental.
\end{enumerate}

As an illustrative demonstration of this pair of existence criteria, consider a Weibull input -- which is characterized by the survival function $\bar{F}\left( t\right) =\exp \left( -\lambda t^{\epsilon }\right) $ (where $\lambda $ and $\epsilon $ are positive parameters). The Weibull distribution \cite {Wei} is one of the three universal laws emanating from the Fisher-Tippett-Gnedenko theorem \cite{FT}-\cite{Gne} of Extreme Value Theory \cite{Gala}-\cite{RT}, and it has numerous uses in science and engineering \cite{MXJ}-\cite{Cool}. This input displays the following scaling property: $\min \left\{T_{1},T_{2}\right\} =2^{-1/\epsilon }T$, where the equality is in law. Consequently, $\mu _{\min }=2^{-1/\epsilon}\mu $, and hence we obtain that: if $\epsilon <1$ then $\mu >2\mu _{\min }$; and if $\epsilon >1$ then $\mu<2\mu _{\min }$. The parameter range $\epsilon <1$ manifests the Stretched Exponential distribution -- which is of major importance in anomalous relaxation phenomena \cite{WW}-\cite{CK1}.

\section{\label{4}Hazard perspective}

At the end of section \ref{3} we noted that the output's mean $M\left( \tau \right) $ -- as a function of the timer parameter $\tau $ -- is flat if and only if the input is Exponential. Hence, given an input $T$ that is not Exponential, it is natural to seek a timer $\tau $ that \emph{minimizes} the output's mean $M\left( \tau \right) $ \emph{globally}. This global minimum can be attained either at the limit $\tau \rightarrow 0$, or at the limit $\tau \rightarrow \infty $, or at a local minimum of the output's mean $M\left( \tau \right)$ (if such exist).

In this section we examine these three global-minimum cases. To that end we employ the notion of hazard function (described below), and use the shorthand notation $\varphi \left( 0\right) =\lim_{t\rightarrow0}\varphi \left( t\right) $ and $\varphi \left( \infty \right) =\lim_{t\rightarrow \infty }\varphi \left( t\right)$ to denote the limit values of a general positive-valued function $\varphi \left( t\right) $ that is defined over the positive half-line ($t>0$); these limit values are assumed to exist in the wide sense, $0\leq \varphi \left( 0\right) ,\varphi \left( \infty \right) \leq \infty $.

The input's \emph{hazard function} is given by the following limit:
\begin{equation}
H\left( t\right) =\lim_{\Delta \rightarrow 0}\frac{1}{\Delta }\Pr \left(T\leq t+\Delta | T\geq t\right) =\frac{f\left(t\right) }{\bar{F}\left( t\right) } \label{41}
\end{equation}
($t>0$). Namely, $H\left( t\right) $ is the likelihood that the input $T$ will be realized right after time $t$, given the information that it was not realized up to time $t$. The hazard function -- also known as \textquotedblleft hazard rate\textquotedblright\ and \textquotedblleft failure rate\textquotedblright\ -- is a widely applied tool in survival analysis \cite{KP}-\cite{Col} and in reliability engineering \cite{BP2,Fin,Dhi}. Now, with the input's hazard function $H\left( t\right) $ at hand, we are all set to examine the global minimum of the output's mean $M\left( \tau \right) $. The derivations of Eqs. (\ref{42})-(\ref{44}) below are detailed in the Methods.

Firstly, we address the case of \emph{fast timers}: $\tau\ll1$. Taking the limit $\tau \rightarrow 0$ in the middle part of Eq. (\ref{32}), while using L'Hospital's rule, yields the limit value 
\begin{equation} M\left( 0\right) =\frac{1}{H\left( 0\right) }.  \label{42}
\end{equation} 
Consequently, we obtain the following pair of \emph{fast criteria}:\footnote{Note that $H(0)=f(0)$, and hence an equivalent formulation of the fast criteria is: $f\left( 0\right) < 1/\mu $ and $f\left( 0\right) > 1/\mu $, respectively.} 

\begin{enumerate}
\item[$\bullet $] If $H\left( 0\right) <1/\mu $ then sharp restart with fast timers is detrimental; in particular, this criterion applies whenever the hazard function vanishes at the origin, $H\left( 0\right) =0$.

\item[$\bullet $] If $H\left( 0\right) >1/\mu $ then sharp restart with fast timers is beneficial; in particular, this criterion applies whenever the hazard function explodes at the origin, $H\left( 0\right) =\infty $.
\end{enumerate}

Secondly, we address the case of \emph{slow timers}: $\tau \gg1$. Taking the limit $\tau \rightarrow \infty $ in Eq. (\ref{32}) yields the limit value $M\left( \infty \right) =\mu $.\footnote{Indeed, an infinite timer $\tau = \infty $ is tantamount to no resetting.} In turn, a calculation using Eq. (\ref{32}) and L'Hospital's rule asserts that the asymptotic behavior of the output's mean $M\left( \tau \right) $ about its limit value $M\left( \infty \right) =\mu $ is given by the following limit: 
\begin{equation}
\lim_{\tau \rightarrow \infty }\frac{M\left( \tau \right) -\mu }{\bar{F}\left( \tau \right) }=\mu -\frac{1}{H\left( \infty \right) }.
\label{43}
\end{equation}%
Consequently, we obtain the following pair of \emph{slow criteria}:

\begin{enumerate}
\item[$\bullet $] If $H\left( \infty \right) <1/\mu $ then sharp restart with slow timers is beneficial; in particular, this criterion applies whenever the hazard function vanishes at infinity, $H\left( \infty \right)
=0 $.

\item[$\bullet $] If $H\left( \infty \right) >1/\mu $ then sharp restart with slow timers is detrimental; in particular, this criterion applies whenever the hazard function explodes at infinity, $H\left( \infty \right)=\infty $.
\end{enumerate}

Thirdly, we address the case of \emph{critical timers}: $\tau _{c}$ for which the derivative of the output's mean vanishes, $M^{\prime }\left( \tau _{c}\right) =0$. Evidently, a local minimum of the output's mean $M\left(\tau _{c}\right) $, if such exists, is attained at a critical timer $\tau_{c} $. A calculation using Eq. (\ref{32}) asserts that, at critical timers (if such exist), the value of the output's mean is given by:
\begin{equation}
M\left( \tau _{c}\right) =\frac{1}{H\left( \tau _{c}\right) }.
\label{44}
\end{equation}
Consequently, we obtain the following pair of \emph{critical criteria}:

\begin{enumerate}
\item[$\bullet $] If $H\left( \tau _{c}\right) <1/\mu $ then sharp restart with the timer $\tau _{c}$ is detrimental.

\item[$\bullet $] If $H\left( \tau _{c}\right) >1/\mu $ then sharp restart with the timer $\tau _{c}$ is beneficial.
\end{enumerate}

\begin{figure}[h!]
\centering
\includegraphics[width=8cm]{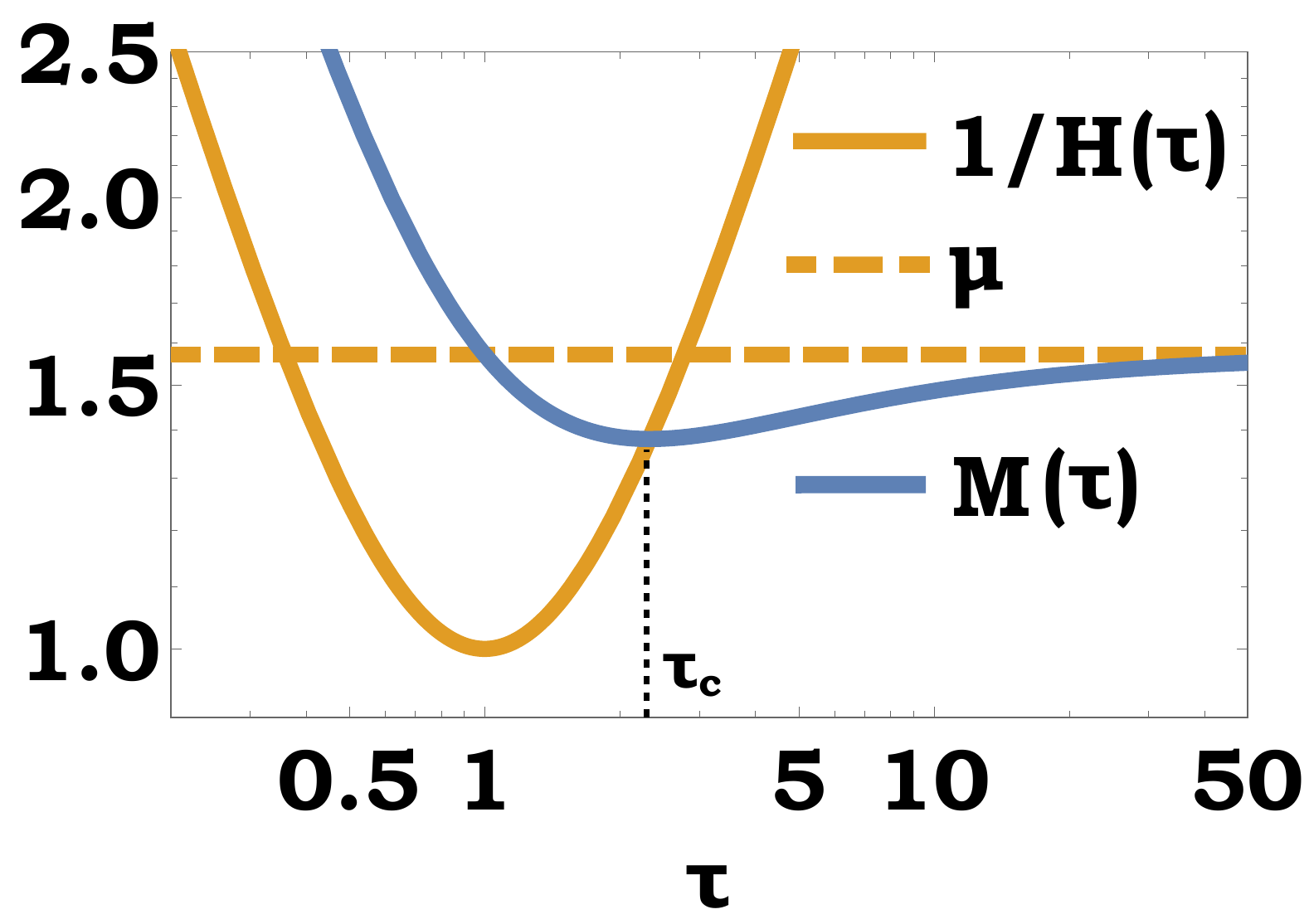}
\caption{Illustration of the results for the fast, slow, and critical timers. We continue with the type-III Pareto input of Fig. 1, for which $\mu = \pi /2 $. The reciprocal $1/H(\tau)=(1+t^2)/(2t)$ of the input's hazard function, and the output’s mean $M(\tau)$, are plotted vs. the timer parameter $\tau$; the $\tau$ axis is depicted on a logarithmic scale. In accord with the fast criteria: for fast timers ($\tau \ll1$) the reciprocal hazard function is larger that input's mean ($1/H(\tau)>\mu$), and sharp restart is detrimental ($M(\tau)>\mu$). In accord with the slow criteria: for slow timers ($\tau \gg1$) the reciprocal hazard function is larger that input's mean ($1/H(\tau)>\mu$), and sharp restart is beneficial ($M(\tau)<\mu$). In accord with the critical criteria: at the critical timer -- the timer $\tau _{c}$ at which the output’s mean attains its global minimum -- the output’s mean and the reciprocal hazard function intersect ($M(\tau _{c})=1/H(\tau _{c})$); also, the reciprocal hazard function is smaller than the input's mean ($1/H(\tau _{c})<\mu$), and sharp restart is beneficial ($M(\tau)<\mu$).}
\label{Hazard_fig2}
\end{figure}

Interestingly, the three pairs of criteria established in this section -- for fast timers ($\tau \ll1$), for slow timers ($\tau \gg1$), and for critical timers ($M^{\prime }\left( \tau _{c}\right) =0$) -- all share a common pattern: comparing the input's hazard function $H(\tau)$ to the level $1/\mu$, the reciprocal of the input's mean. An equivalent formulation of the common pattern is: comparing the reciprocal $1/H(\tau)$ of the input's hazard function to the input's mean $\mu$ (Fig. 2).

The reciprocal of the input's mean, $1/\mu$, can be interpreted as an average ``task-completion rate''. With this interpretation in mind, the fast and slow restart criteria admit the following intuitive explanations. If $H(0)>1/\mu$ then the rate at times $t \ll1$ is larger than the average rate; consequently, restart with fast timers increases the task-completion rate (by including only times $t \ll1$), and is hence beneficial. If $H(\infty)<1/\mu$ then the rate at times $t \gg 1$ is smaller than the average rate; consequently, restart with slow timers increases the task-completion rate (by excluding times $t \gg1$), and is hence beneficial. The intuitive explanations for detrimental restart with fast and slow timers are analogous. These intuitive explanations elucidate why the inequalities appearing in the fast criteria are opposite to the inequalities appearing in the slow criteria.

The flipping of the inequalities can also be understood by examining the arrow of time. Indeed, for a fast timer $\tau$, the fast criteria check $H(0)$ -- the value of the hazard function at the origin, i.e. \emph{before} the timer. And, for a critical timer $\tau_{c}$, the critical criteria check $H(\tau _{c})$ -- the value of the hazard function \emph{at} the timer. On the other hand, for a slow timer $\tau$, the slow criteria check $H(\infty)$ -- the value of the hazard function at infinity, i.e. \emph{after} the timer. Namely, with respect to the arrow of time: the fast criteria check a `past value' of the hazard function, the critical criteria check a `present value', and the slow criteria check a `future value'. The flipping of the inequalities manifests the transition from the (observed) past and present to the (unobserved) future.

In the next two sections we shall further explore the cases of fast, slow, and critical timers. In particular: section \ref{5} shall address the ranges of fast and slow timers; and section \ref{6} shall address time-reversal, i.e. the reversal of the arrow of time.

\section{\label{5}Fast and slow timers}

In the previous section we addressed the case of \emph{fast timers} ($\tau\ll1$), and the case of \emph{slow timers} ($\tau \gg1$). However, we did not provide the precise meanings of these timers. Namely: how small should the timer $\tau $ be in order to qualify as `fast'? and how large should the timer $\tau $ be in order to qualify as `slow'? Considering the input's hazard function $H\left( t\right) $ to be continuous over the positive half-line ($t>0$), in this section we answer these two questions.

The interplay between the following terms assumed a key role in the previous section: the input's hazard function $H\left( t\right) $ on the one hand, and the input's mean $\mu $ on the other hand. This interplay will assume a key role also in this section -- via the two following thresholds:
\begin{equation}
\tau _{\ast }=\inf \left\{ t>0 \:|\: H\left( t\right) =%
\frac{1}{\mu }\right\} ,  \label{51}
\end{equation}
and
\begin{equation}
\tau ^{\ast }=\sup \left\{ t>0 \:|\: H\left( t\right) =%
\frac{1}{\mu }\right\} .  \label{52}
\end{equation}
Namely, the lower threshold $\tau _{\ast }$ and the upper threshold $\tau ^{\ast }$ are, respectively, the smallest and largest times at which the hazard function $H\left( t\right)$ intersects the level $1/\mu$ (Fig. 3). In particular, if the hazard function $H\left( t\right) $ does \emph{not} intersect the level $1/\mu $ then: $\tau _{\ast }=\infty $ and $\tau ^{\ast }=0$.

Combining together Eq. (\ref{32}) and the input's hazard rate, the difference between the output's mean and the input's mean admits the following formulations:
\begin{equation}
M\left( \tau \right) -\mu =\frac{1}{F\left( \tau \right) }\int_{0}^{\tau }\left[ \frac{1}{H\left( t\right) }-\mu \right] f\left(t\right) dt.  \label{53}
\end{equation}
and
\begin{equation}
M\left( \tau \right) -\mu =\frac{1}{F\left( \tau \right) }\int_{\tau}^{\infty }\left[ \mu -\frac{1}{H\left( t\right) }\right] f\left( t\right) dt.  \label{54}
\end{equation}%
The derivations of Eqs. (\ref{53}) and (\ref{54}) are detailed in the Methods. Observing Eqs. (\ref{53}) and (\ref{54}), it is evident that the sign of the difference $M(\tau) -\mu$ depends on the interplay between the input's hazard function and mean. Armed with the thresholds of Eqs. (\ref{51}) and (\ref{52}), as well as with the mean-difference formulae of Eqs. (\ref{53}) and (\ref{54}), we are all set to analyze the cases of fast and slow timers. 

Consider the case of \emph{fast timers}: $\tau \ll1$. The fast-timer criteria of the previous section, combined together with Eq. (\ref{53}), straightforwardly yields the two following conclusions. (I) If $H\left(0\right) <1/\mu $ then sharp restart is detrimental for all timers $\tau $ in the range $\tau <\tau _{\ast }$. (II) If $H\left( 0\right) >1/\mu $ then sharp restart is beneficial for all timers $\tau $ in the range $\tau <\tau_{\ast }$. Hence, the lower threshold $\tau _{\ast }$ of Eq. (\ref{51}) defines a \emph{range} of \emph{fast timers}. 

Consider the case of \emph{slow timers}: $\tau \gg1$. The slow-timer criteria of the previous section, combined together with Eq. (\ref{54}), straightforwardly yields the two following conclusions. (I) If $H\left(\infty \right) <1/\mu $ then sharp restart is beneficial for all timers $\tau $ in the range $\tau >\tau ^{\ast }$. (II) If $H\left( \infty \right)>1/\mu $ then sharp restart is detrimental for all timers $\tau $ in the range $\tau >\tau ^{\ast }$. Hence, the upper threshold $\tau ^{\ast }$ of Eq. (\ref{52}) defines a \emph{range} of \emph{slow timers}. 

At the end of section \ref{3} we noted the special and unique role of Exponential inputs. Specifically, these are the only inputs that produce a flat output mean: $M\left( \tau \right) =\mu $ for all timers $\tau $. An Exponential input is characterized by the flat hazard function $1/\mu $. Consequently -- for a general input $T$ -- the difference between the reciprocal $1/H\left( t\right) $ of the input's hazard function and the input's mean $\mu$ is actually: the difference between the reciprocals of two hazard functions -- one of the general input, and one of an Exponential input with mean $\mu $. Thus, in a `hazard-function sense', the integrals in Eqs. (\ref{53}) and (\ref{54}) measure the deviation of the general input $T$ from an Exponential input with the same mean ($\mu $).

\begin{figure}[t!]
\centering
\includegraphics[width=8cm]{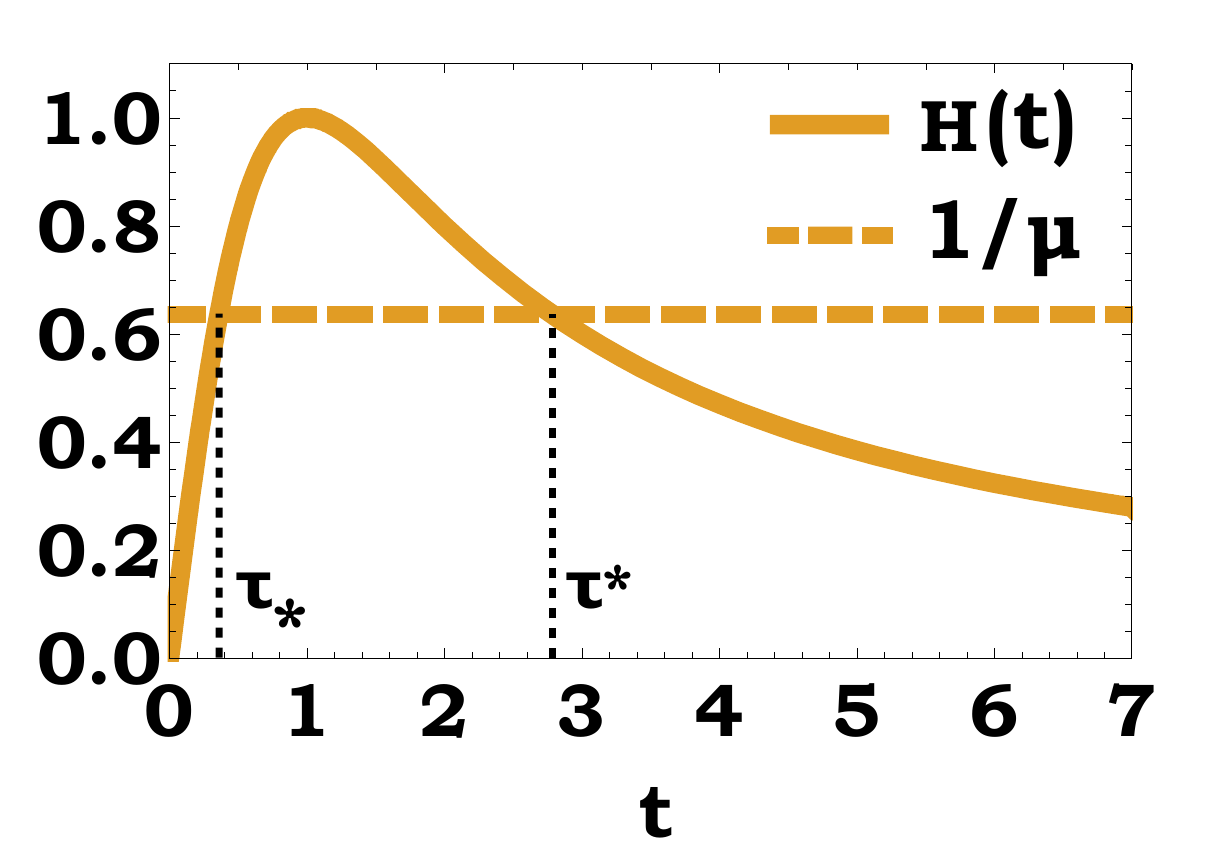}
\caption{Illustration of the lower and upper thresholds, $\tau _{\ast }$ and $\tau ^{\ast }$. We continue with the type-III Pareto input of Fig. 1 and Fig. 2, for which: the mean is $\mu = \pi /2 $, and the hazard function is $H(t)=2t/(1+t^2)$; this mean level and this hazard function are plotted here. In turn, the lower threshold is $\tau _{\ast }=(\pi - \sqrt{\pi^2 - 4})/2 $, and the upper threshold is $\tau ^{\ast } =(\pi + \sqrt{\pi^2 - 4})/2$. In Fig. 1 we saw that sharp restart with timers $\tau<1$ is detrimental, and that sharp restart with timers $\tau>1$ is beneficial. The lower and upper thresholds are in accord with Fig. 1: the range of fast timers ($\tau<\tau _{\ast }$) is included in the range of `detrimental timers' ($\tau<1$), and the range of slow timers ($\tau > \tau ^{\ast }$) is included in the range of `beneficial timers' ($\tau>1$).}
\label{Fast-Slow_fig}
\end{figure}

\section{\label{6}Critical timers}

In section \ref{4} we addressed the case of \emph{critical timers}: $\tau_{c}$ for which the derivative of the output's mean vanishes, $M^{\prime}\left( \tau _{c}\right) =0$. Evidently, these are the timers at which the local minima and the local maxima -- if such exist -- of the output's mean $M\left( \tau \right)$ are attained. In this section we elaborate on critical timers. 

To analyze the critical timers we take the logarithm of the left and middle parts of Eq. (\ref{32}), and differentiate this logarithm twice. As we shall now describe, the first differentiation gives rise to the notion of backward hazard function, and the second differentiation gives rise to the notion of Gibbs gradient function. Using these two functions, we shall now establish a detailed characterization of critical timers.

The hazard function, which we employed quite extensively in sections \ref{4} and \ref{5}, implicitly assumes that the arrow of time points \emph{forward}. But what if the arrow of time is reversed, and time flows \emph{backward} rather than forward? In such a time-reversal setting we shift from the input's `forward' hazard function of Eq. (\ref{41}) to the following `backward' hazard function:

\begin{equation}
B\left( t\right) =\lim_{\Delta \rightarrow 0}\frac{1}{\Delta }\Pr \left(
T>t-\Delta | T\leq t\right) =\frac{f\left( t\right) 
}{F\left( t\right) }. \label{61}
\end{equation}%
Namely, $B\left( t\right) $ is the likelihood that the input $T$ will be
realized right before time $t$, given the information that it was not
realized after time $t$.\footnote{In the forward hazard function of Eq. (\ref{41}) we observe the past -- the time interval $\left( 0,t\right) $; then, based on this observation, we predict the present -- time $t$. Conversely, in the backward hazard function of Eq. (\ref{61}) we observe the future -- the time ray $\left( t,\infty
\right) $; then, based on this observation, we predict the present -- time $t$.}

Interestingly, critical timers link together the notion of residual lifetime (which we described and used in section \ref{3}), and the notion of backward hazard function. Indeed, consider the input's backward hazard function $B\left( t\right)$ of Eq. (\ref{61}). Also, consider the backward hazard function, $B_{res}\left( t\right) $ ($t>0$), of the input's residual lifetime $T_{res}$; this function is defined identically to Eq. (\ref{61}). In terms of these two backward hazard functions, the critical timers $\tau _{c}$ are identified as follows:
\begin{equation}
M^{\prime }\left( \tau _{c}\right) =0 \, \Leftrightarrow \,
B\left( \tau _{c}\right) =B_{res}\left( \tau _{c}\right) .
\label{62}
\end{equation}%
Namely, the critical timers $\tau _{c}$ (if such exist) are the intersection points (if such exist) of the two backward hazard functions: that of the input, $B\left( t\right) $; and that of the input's residual lifetime, $B_{res}\left( t\right) $. The derivation of Eq. (\ref{62}) -- via the differentiation of the logarithm of Eq. (\ref{32}) -- is detailed in the Methods.

Considering the input's density function $f\left( t\right) $ to be smooth, we introduce its negative logarithmic derivative:
\begin{equation}
G\left( t\right) =-\frac{d}{dt}\ln \left[ f\left( t\right) \right] =-\frac{%
f^{\prime }\left( t\right) }{f\left( t\right) },  \label{63}
\end{equation}
($t>0$). The negative logarithmic derivative $G\left( t\right) $ has a profound meaning: up to a scale factor, it is the gradient of the potential function that underpins the \emph{Gibbs representation} of the input's density \cite{Geo}-\cite{AvOv}. The Gibbs representation emerges via entropy maximization \cite{Jay1}-\cite{Kap}, as well as via the steady-state statistics of the Langevin equation \cite{Lan}-\cite{Pav}.

Consider the input's Gibbs gradient function $G\left( t\right) $ of Eq. (\ref{63}). Also, consider the Gibbs gradient function, $G_{res}\left( t\right) $($t>0$), of the input's residual lifetime $T_{res}$; this function is defined identically to Eq. (\ref{63}). In terms of these two Gibbs gradient functions, the behavior at critical timers $\tau _{c}$ is determined as follows:

\begin{enumerate}
\item[$\bullet $] If $G\left( \tau _{c}\right) <G_{res}\left( \tau _{c}\right) $ then the critical timer $\tau _{c}$ yields a local maximum of the output's mean $M\left( \tau \right) $. 

\item[$\bullet $] If $G\left( \tau _{c}\right) >G_{res}\left( \tau _{c}\right) $ then the critical timer $\tau _{c}$ yields a local minimum of the output's mean $M\left( \tau \right) $. 
\end{enumerate}

\noindent The derivation of these two results -- via the double differentiation of the logarithm of Eq. (\ref{32}) -- is detailed in the Methods.

The results presented in this section are analogues to the residual criteria of section \ref{3}. One the one hand, for a general timer $\tau$, the residual criteria of section \ref{3} compare the survival function of the input $T$ to that of the input's residual lifetime $T_{res}$. On the other hand, for a critical timer $\tau_{c}$, the results of this section compare the backward hazard function and the Gibbs gradient function of the input $T$ to that of the input's residual lifetime $T_{res}$. These comparisons further expose the profound relation between sharp restart and the residual lifetime.

\section{\label{7} Summary and discussion}

This paper presented a comprehensive, statistical based, mean-performance analysis of the \emph{sharp-restart algorithm}. The algorithm takes as input the random completion time $T$ of a general stochastic process. This process is restarted  periodically -- using a sharp (deterministic) timer $\tau $ -- until it reaches completion. The algorithm produces as output a new random completion time $T_{R}$, the overall time it takes to perform the task under sharp restart. The analysis focused on comparing the input's mean $\mathbf{E}\left[ T\right]=\mu$ to the output's mean $\mathbf{E}\left[ T_R\right]=M\left( \tau \right) $ (which is a function of the timer $\tau $). The two principal analytic tools employed were the \emph{residual lifetime}\ of renewal theory, and the \emph{hazard rate}\ of reliability engineering. Using these tools, we established a detailed \emph{statistical} \emph{roadmap} for the mean-performance of sharp-restart: six pairs of universal criteria that determine if the application of sharp restart is beneficial $M\left( \tau \right) <\mu $, or detrimental $M\left(\tau \right) >\mu $. The criteria are summarized in Table 1.

\begin{table}[]\label{table1}
\begin{center}
{\LARGE Table 1}

\ \ 

\begin{tabular}{||l||l||l||l||l||}
\hline\hline
& $%
\begin{array}{c}
\  \\ 
\textbf{Timer} \\ 
\ \\
\end{array}%
$ & $%
\begin{array}{c}
\ \\ 
\textbf{Parameter} \\ 
\
\end{array}%
$ & $%
\begin{array}{c}
\ \\ 
\quad \textbf{Beneficial}\\ 
\ \\
\end{array}%
$ & $%
\begin{array}{c}
\ \\ 
\enspace \textbf{Detrimental} \\ 
\
\end{array}%
$ \\ \hline\hline
\textbf{I} & $%
\begin{array}{c}
\ \\ 
General \\ 
\ \\
\end{array}%
$ & $0<\tau <\infty $ & $\bar{F}\left( \tau \right) <\bar{F}_{res}\left(
\tau \right) $ & $\bar{F}\left( \tau \right) >\bar{F}_{res}\left( \tau
\right) $ \\ \hline\hline
\textbf{II} & $%
\begin{array}{c}
\ \\ 
Existence \\ 
\ \\
\end{array}%
$ & \qquad --- & $%
\begin{array}{c}
\  \\
\enspace \quad \mu <\mu _{res}\\
\enspace \quad \Updownarrow  \\
\enspace \quad \mu <\sigma \\
\ 
\end{array}%
$ & $%
\begin{array}{c}
\ \\
\enspace \quad \mu >\mu _{res}\\  
\enspace \quad \Updownarrow \\  
\enspace \quad \mu >\sigma \\ 
\ 
\end{array}%
$ \\ \hline\hline
\textbf{III} & $%
\begin{array}{c}
\ \\ 
Existence \\ 
\ 
\end{array}%
$ & \qquad --- & $%
\begin{array}{c}
\ \\ 
\mu >2\mu _{\min } \\ 
\Updownarrow \\ 
\Pr \left( T>T_{res}\right) <\frac{1}{2} \\ 
\ 
\end{array}%
$ & $%
\begin{array}{c}
\ \\ 
\mu <2\mu _{\min } \\ 
\Updownarrow \\ 
\Pr \left( T>T_{res}\right) >\frac{1}{2} \\ 
\ 
\end{array}%
$ \\ \hline\hline
\textbf{IV} & $%
\begin{array}{c}
\ \\ 
Fast \\ 
\ 
\end{array}%
$ & $0< \tau < \tau _{\ast }$ & $H\left( 0\right) >\frac{1}{\mu }$ & $H\left(
0\right) <\frac{1}{\mu }$ \\ \hline\hline
\textbf{V} & $%
\begin{array}{c}
\ \\ 
Critical \\ 
\
\end{array}%
$ & $M^{\prime }\left( \tau _{c}\right) =0$ & $H\left( \tau _{c}\right) >%
\frac{1}{\mu }$ & $H\left( \tau _{c}\right) <\frac{1}{\mu }$ \\ \hline\hline
\textbf{VI} & $%
\begin{array}{c}
\ \\ 
Slow \\ 
\
\end{array}%
$ & $\tau^{\ast } < \tau < \infty$ & $H\left( \infty \right) <\frac{1}{\mu }$
& $H\left( \infty \right) >\frac{1}{\mu }$ \\ \hline\hline
\end{tabular}
\end{center}

\small{\textbf{Table 1}: Six pairs of universal mean-performance criteria for the sharp-restart algorithm. The Table's columns specify the features of each pair of criteria: to which timer parameters $\tau $ do the criteria apply; and when is the application of the algorithm beneficial/detrimental.  \textbf{I}) For a general timer $\tau $ (section \ref{3}): the criteria compare -- at the point $\tau $ -- the value of the input's survival function, $\bar{F}\left( \tau \right) $, to the value of the survival function of the input's residual lifetime, $\bar{F}_{res}\left(\tau \right) $. \textbf{II}) To determine the existence of timers for which sharp-restart is beneficial/detrimental (section \ref{8}): the criteria compare the input's mean, $\mu $, to the mean of the input's residual lifetime, $\mu _{res}$; equivalently, the criteria compare the input's mean, $\mu $, to the input's standard deviation, $\sigma $. \textbf{III}) To determine the existence of timers for which sharp-restart is beneficial/detrimental (section \ref{8}): the criteria compare the input's mean, $\mu $, to twice the mean of the minimum of two IID copies of the input, $2 \mu _{min}$; equivalently, the criteria examine the probability that the input $T$ be greater than the input's residual lifetime $T_{res}$ (where the random variables $T$ and $T_{res}$ are independent of each other). \textbf{IV}) For fast timers $0<\tau<\tau_{*}$ (sections \ref{4} and \ref{5}): the criteria compare the value of input's hazard function at zero, $H(0)$, to the reciprocal of the input's mean, $1/\mu $. \textbf{V}) For a critical timer $\tau_{c} $ (sections \ref{4} and \ref{6}): the criteria compare the value of input's hazard function at the point $\tau_{c} $,  $H(\tau_{c}) $, to the reciprocal of the input's mean, $1/\mu $. \textbf{VI}) For slow timers $\tau^{*}<\tau< \infty$ (sections \ref{4} and \ref{5}): the criteria compare the value of input's hazard function at infinity, $H(\infty)$, to the reciprocal of the input's mean, $1/\mu $. }
\end{table}

As stated in the introduction, the most common and widely applied restart protocol in the physics literature is exponential restart -- which uses exponentially-distributed timers. To appreciate the six pairs of universal sharp-restart criteria that were established here, one has to compare Table 1 to a corresponding exponential-restart table. To date, rows \textbf{I-V} in the exponential-restart table are missing; indeed, the exponential-restart criteria that should appear in these rows are not available. 

As noted in section \ref{8} above, row \textbf{VI} in the exponential-restart table is known. Namely, with regard to exponentially-distributed timers with large means, it was shown that \cite{FPUR0,FPUR1,HRS,Restart-Mol1,Optimal_fluc}: if $\sigma > \mu$ then exponential restart is beneficial; and if $\sigma < \mu$ then exponential restart is detrimental. Comparing this result to row \textbf{VI} of Table 1 highlights a profound difference between exponential restart and sharp restart, which arises in the following scenario:
\begin{equation}
\sigma < \mu < \frac{1}{H(\infty)}.
\label{81}
\end{equation}
Namely, for inputs meeting the scenario of Eq. (\ref{81}), exponential restart with slow timers is detrimental, whereas sharp restart with slow timers is beneficial.

We shall demonstrate the scenario of Eq. (\ref{81}) via two examples of canonical diffusion processes: diffusions in linear and logarithmic potentials. To describe these examples, consider a diffusion process \cite{VK} that runs over the non-negative half-line. The stochastic dynamics of the diffusion process are governed by a Langevin equation \cite{CK} with potential function  $U(x)$ ($0 \leq x< \infty$) and with positive drag and diffusion coefficients, $\zeta$ and $D$, respectively. Namely, the Langevin equation is:
\begin{equation}
\dot{X}(t)=- \frac{1}{\zeta} \cdot U'[X(t)]+\sqrt{2D} \cdot \eta(t),
\end{equation}
where $\eta(t)$ is a Gaussian white noise.\footnote{The Gaussian white noise has zero mean $\langle\eta(t)\rangle=0$, and Dirac delta-function auto-correlation $\langle\eta(t)\eta(t')\rangle=\delta(t-t')$.} Initiating the diffusion process from the positive level $l$, we set the random variable $T$ to be the first-passage time to the origin, i.e. $T$ is the first time the diffusion process reaches the level $0$.

The linear-potential example is given by \cite{Red}: $U(x)=U_1 x$, where $U_1$ is a positive slope. The resulting diffusion process is Brownian motion with negative drift in an average `speed' $v=U_1/\zeta$. The resulting first-passage time $T$ is an inverse-Gauss random variable with the following density function \cite{Red}:
\begin{equation}
f(t)=\frac{l}{\sqrt{4 \pi Dt^3}} \exp{\left[-\frac{(l-vt)^2}{4Dt}\right]}.
\end{equation}
This inverse-Gauss first-passage time has: mean $\mu=l/v$; squared coefficient of variation $\sigma^2 / \mu^2 = 2D/(lv)$; and hazard limit $H(\infty)=v^2/(4D)$.  Consequently, in terms of the P\'eclet number $Pe=lv/(2D)$ -- which manifests the ratio between the underlying rates of drift and diffusion \cite{Red} -- the scenario of Eq. (\ref{81}) holds in the rage $1<Pe<2$. With regard to slow timers, we thus obtain that: in the passage from exponential restart to sharp restart the P\'eclet transition point jumps from $Pe=1$ \cite{FPUR5,PotD1} to $Pe=2$.

The logarithmic-potential example is given by \cite{AJBray}-\cite{log2}: $U(x)=U_e\log(x)$, where $U_e$ is a real constant. Recall that the Einstein relation couples the drag and diffusion coefficients via the ``thermodinamic beta'', $\beta=1/(\zeta D)$. Now, provided that $\beta U_e > -1$, the resulting first-passage time $T$ is an inverse-Gamma random variable with the following density function \cite{AJBray,ekaterina}:
\begin{equation}
f(t)= \frac{4D}{\Gamma{(\nu)l^2}} \left(\frac{l^2}{4Dt} \right)^{\nu+1} \exp{\left(-\frac{l^2}{4Dt}\right)},
\end{equation}
where $\nu=(\beta U_e+ 1)/2$. This inverse-Gamma first-passage time has: mean $\mu=l^2/[2D(\beta U_e-1)]$ (provided that $\beta U_e>1$); squared coiffcient of variation $\sigma^2 / \mu^2 = 2/(\beta U_e-3)$ (provided that $\beta U_e>3$); and hazard limit $H(\infty)=0$. Consequently, the scenario of Eq. (\ref{81}) holds in the parameter range $\beta U_e>5$. With regard to slow timers, we thus obtain that: while exponential restart is beneficial only when $\beta U_e<5$ \cite{PotD3}, sharp restart is always beneficial.

The above inverse-Gamma density function displays a power-law asymptotic decay -- which, in turn, is a particular example of a density function whose corresponding hazard function vanishes at infinity, $H(\infty)=0$. Stretched-Exponential first-passage times -- which appear prevalently in anomalous relaxation \cite{WW}-\cite{CK1}, and which were noted in subsection \ref{Res} above -- also have $H(\infty)=0$. So do Lognormal completion times \cite{Gal}-\cite{Dow}, which are observed as service times in call centers \cite{BGM}-\cite{GT2}. In all these examples, according to row \textbf{VI} of Table 1, sharp restart with slow timers is always beneficial. Moreover, if $H(\infty)=0$ then the scenario of Eq. (\ref{81}) simplifies to $\sigma < \mu < \infty$. We thus see that the scenario of Eq. (\ref{81}) broadly applies to first-passage/completion times that: on the one hand, have a relatively small standard deviation ($\sigma/\mu < 1$); and, on the other hand, have slowly decaying probability tails ($H(\infty)=0$).

We conclude with a brief outlook. This paper shall be followed by a sequel that explores the mean-performance of sharp restart via a comprehensive socioeconomic-inequality analysis \cite{Socioeconomic}. Tail-performance of sharp restart offers a complementary approach to mean-performance, which will be explored elsewhere \cite{Tail-behavior}. Alternative to the mean, other performance measures -- e.g., median and mode -- have been investigated for exponential restart \cite{FPUR9}; it is of interest to extend such investigations to sharp restart. This paper implicitly assumed that restart is instantaneous, i.e., that the resetting of the underlying process takes zero time. As in various systems restart is non-instantaneous \cite{FPUR0,FPUR3,FPUR6,FPUR7,FPUR8,HRS,Restart-Mol1,Restart-Mol5,Restart-Mol6,freeD4}, it is also of interest to study the effect of sharp restart on such systems. Indeed, there are many open restart-research challenges awaiting to be addressed in the coming future. \\

\textbf{Acknowledgments}. Shlomi Reuveni acknowledges support from the Azrieli Foundation, from the Raymond and Beverly Sackler Center for Computational Molecular and Materials Science at Tel Aviv University, and from the Israel Science Foundation (grant No. 394/19). 

\newpage

\section{Methods}

\subsection{Derivation of Eq. (\protect\ref{22})}

Applying expectation to both sides of Eq. (\ref{21}), while using the
properties of the sharp-restart algorithm, yields%
\begin{equation}
\begin{array}{l}
\mathbf{E}\left[ T_{R}\right] =\mathbf{E}\left[ I\left\{ T\leq \tau \right\}
\cdot T+I\left\{ T>\tau \right\} \cdot \left( \tau +T_{R}^{\prime }\right) %
\right] \\ 
\ \\ 
=\mathbf{E}\left[ I\left\{ T\leq \tau \right\} \cdot T\right] +\mathbf{E}%
\left[ I\left\{ T>\tau \right\} \cdot \left( \tau +T_{R}^{\prime }\right) %
\right] \\ 
\ \\ 
=\int_{0}^{\infty }\left[ I\left\{ t\leq \tau \right\} t\right] f\left(
t\right) dt+\mathbf{E}\left[ I\left\{ T>\tau \right\} \right] \cdot \mathbf{E%
}\left[ \tau +T_{R}^{\prime }\right] \\ 
\ \\ 
=\int_{0}^{\tau }tf\left( t\right) dt+\Pr \left( T>\tau \right) \cdot \left(
\tau +\mathbf{E}\left[ T_{R}^{\prime }\right] \right) \\ 
\ \\ 
=\int_{0}^{\tau }tf\left( t\right) dt+\bar{F}\left( \tau \right) \cdot
\left( \tau +\mathbf{E}\left[ T_{R}\right] \right) .%
\end{array}
\label{A32}
\end{equation}%
Note that%
\begin{equation}
\left. 
\begin{array}{l}
\int_{0}^{\tau }tf\left( t\right) dt=\int_{0}^{\tau }\left[ \int_{0}^{t}du%
\right] f\left( t\right) dt \\ 
\ \\ 
=\int_{0}^{\tau }\left[ \int_{u}^{\tau }f\left( t\right) dt\right]
du=\int_{0}^{\tau }\left[ \bar{F}\left( u\right) -\bar{F}\left( \tau \right) %
\right] du \\ 
\ \\ 
=\int_{0}^{\tau }\bar{F}\left( u\right) du-\tau \bar{F}\left( \tau \right) 
.%
\end{array}%
\right.  \label{A33}
\end{equation}%
Substituting Eq. (\ref{A33}) into Eq. (\ref{A32}) yields%
\begin{equation}
\mathbf{E}\left[ T_{R}\right] =\int_{0}^{\tau }\bar{F}\left( u\right) du+%
\bar{F}\left( \tau \right) \cdot \mathbf{E}\left[ T_{R}\right] .
\label{A34}
\end{equation}%
In turn, Eq. (\ref{A34}) implies Eq. (\ref{22}):%
\begin{equation}
M\left( \tau \right)=\mathbf{E}\left[ T_{R}\right] =\frac{\int_{0}^{\tau }\bar{F}\left( u\right)
du}{1-\bar{F}\left( \tau \right) }=\frac{1}{F\left( \tau \right) }%
\int_{0}^{\tau }\bar{F}\left( u\right) du.  \label{A35}
\end{equation}

\subsection{Derivation of Eqs. (\protect\ref{36})-(\protect\ref{37})}

Eq. (\ref{32}) implies that%
\begin{equation}
\frac{M\left( \tau \right) -\mu }{\mu }=\frac{F_{res}\left( \tau \right)
-F\left( \tau \right) }{F\left( \tau \right) }=\frac{\bar{F}\left( \tau
\right) -\bar{F}_{res}\left( \tau \right) }{F\left( \tau \right) }.
\label{A80}
\end{equation}%
In what follows the random variables $T_{1}$ and $T_{2}$ are IID copies of
the input $T$.

The distribution function of the maximum $T_{\max }=\max \left\{
T_{1},T_{2}\right\} $ is: $\Pr \left( T_{\max }\leq t\right) =F\left(
t\right) ^{2}$ ($t\geq 0$). Consequently, the maximum's density function is $%
f_{\max }\left( t\right) =2F\left( t\right) f\left( t\right) $ ($t>0$). In
turn, Eq. (\ref{A80}) implies that 
\begin{equation}
\left. 
\begin{array}{l}
\frac{M\left( \tau \right) -\mu }{\mu }f_{\max }\left( \tau \right) =\left[
F_{res}\left( \tau \right) -F\left( \tau \right) \right] 2f\left( \tau
\right) \\ 
\ \\ 
=2F_{res}\left( \tau \right) f\left( \tau \right) -f_{\max }\left( \tau
\right) .%
\end{array}%
\right.  \label{A83}
\end{equation}%
Integrating both sides of Eq. (\ref{A83}) over the positive half-line yields%
\begin{equation}
\left. 
\begin{array}{l}
\int_{0}^{\infty }\left[ \frac{M\left( \tau \right) -\mu }{\mu }\right]
f_{\max }\left( \tau \right) d\tau \\ 
\ \\ 
=\int_{0}^{\infty }\left[ 2F_{res}\left( \tau \right) f\left( \tau \right)
-f_{\max }\left( \tau \right) \right] d\tau \\ 
\ \\ 
=2\int_{0}^{\infty }F_{res}\left( \tau \right) f\left( \tau \right) d\tau
-\int_{0}^{\infty }f_{\max }\left( \tau \right) d\tau \\ 
\ \\ 
=2\int_{0}^{\infty }F_{res}\left( \tau \right) f\left( \tau \right) d\tau -1%
.%
\end{array}%
\right.  \label{A84}
\end{equation}

In what follows the random variables $T$ and $T_{res}$ are considered
independent of each other. Note that%
\begin{equation}
\left. 
\begin{array}{l}
\Pr \left( T>T_{res}\right) =\Pr \left( T_{res}\leq T\right) \\ 
\ \\ 
=\int_{0}^{\infty }\Pr \left( T_{res}\leq t | 
T=t\right) f\left( t\right) dt \\ 
\ \\ 
=\int_{0}^{\infty }F_{res}\left( t\right) f\left( t\right) dt.%
\end{array}%
\right.  \label{A85}
\end{equation}%
Substituting Eq. (\ref{A85}) into Eq. (\ref{A84}) yields Eq. (\ref{36}):%
\begin{equation}
\int_{0}^{\infty }\left[ \frac{M\left( \tau \right) -\mu }{\mu }\right]
f_{\max }\left( \tau \right) d\tau =2\Pr \left( T>T_{res}\right) -1.
\label{A81}
\end{equation}

The survival function of the minimum $T_{\min }=\min \left\{
T_{1},T_{2}\right\} $ is: $\Pr \left( T_{\min }>t\right) =\bar{F}\left(
t\right) ^{2}$ ($t\geq 0$). Consequently, the minimum's mean is $\mu _{\min
}=\int_{0}^{\infty }\bar{F}\left( t\right) ^{2}dt$. In turn, using Eq. (\ref%
{31}), we have%
\begin{equation}
\left. 
\begin{array}{l}
\Pr \left( T>T_{res}\right) =\int_{0}^{\infty }\Pr \left( T>t | T_{res}=t\right) f_{res}\left( t\right) dt \\ 
\ \\ 
=\int_{0}^{\infty }\bar{F}\left( t\right) f_{res}\left( t\right)
dt=\int_{0}^{\infty }\bar{F}\left( t\right) \left[ \frac{1}{\mu }\bar{F}%
\left( t\right) \right] dt \\ 
\ \\ 
=\frac{1}{\mu }\int_{0}^{\infty }\bar{F}\left( t\right) ^{2}dt=\frac{\mu
_{\min }}{\mu }.%
\end{array}%
\right.  \label{A87}
\end{equation}%
Substituting Eq. (\ref{A87}) into Eq. (\ref{A81}) yields%
\begin{equation}
\int_{0}^{\infty }\left[ \frac{M\left( \tau \right) -\mu }{\mu }\right]
f_{\max }\left( \tau \right) d\tau =\frac{2\mu _{\min }-\mu }{\mu }.
\label{A88}
\end{equation}%
In turn, Eq. (\ref{A88}) yields Eq. (\ref{37}).

\subsection{Derivation of Eqs. (\protect\ref{42})-(\protect\ref{44})}

Eqs. (\ref{31}) and (\ref{41}) imply that
\begin{equation}
\frac{f\left( t\right) }{f_{res}\left( t\right) }=\frac{f\left( t\right) }{%
\frac{1}{\mu }\bar{F}\left( t\right) }=\mu H\left( t\right) .  \label{A40}
\end{equation}%
Taking the limit $\tau \rightarrow 0$ in the middle part of Eq. (\ref{32}),
and using L'Hospital's rule and Eq. (\ref{A40}), yields%
\begin{equation}
\left. 
\begin{array}{l}
\frac{M\left( 0\right) }{\mu }=\lim_{\tau \rightarrow 0}\frac{M\left( \tau
\right) }{\mu } \\ 
\ \\ 
=\lim_{\tau \rightarrow 0}\frac{F_{res}\left( \tau \right) }{F\left( \tau
\right) }=\lim_{\tau \rightarrow 0}\frac{f_{res}\left( \tau \right) }{%
f\left( \tau \right) } \\ 
\ \\ 
=\lim_{\tau \rightarrow 0}\frac{1}{\mu H\left( \tau \right) }=\frac{1}{\mu
H\left( 0\right) }.%
\end{array}%
\right.  \label{A42}
\end{equation}%
In turn, Eq. (\ref{A42}) yields Eq. (\ref{42}).

Eq. (\ref{A80}) implies that%
\begin{equation}
\frac{M\left( \tau \right) -\mu }{\mu \bar{F}\left( \tau \right) }=\frac{%
\bar{F}\left( \tau \right) -\bar{F}_{res}\left( \tau \right) }{F\left( \tau
\right) \bar{F}\left( \tau \right) }=\frac{1}{F\left( \tau \right) }\left[ 1-%
\frac{\bar{F}_{res}\left( \tau \right) }{\bar{F}\left( \tau \right) }\right] 
.  \label{A43}
\end{equation}%
Taking the limit $\tau \rightarrow \infty $ in the right part of Eq. (\ref%
{A43}), and using L'Hospital's rule and Eq. (\ref{A40}), yields%
\begin{equation}
\left. 
\begin{array}{l}
\lim_{\tau \rightarrow \infty }\frac{M\left( \tau \right) -\mu }{\mu \bar{F}%
\left( \tau \right) }=\lim_{\tau \rightarrow \infty }\frac{1}{F\left( \tau
\right) }\left[ 1-\frac{\bar{F}_{res}\left( \tau \right) }{\bar{F}\left(
\tau \right) }\right] \\ 
\ \\ 
=1-\lim_{\tau \rightarrow \infty }\frac{\bar{F}_{res}\left( \tau \right) }{%
\bar{F}\left( \tau \right) }=1-\lim_{\tau \rightarrow \infty }\frac{%
f_{res}\left( \tau \right) }{f\left( \tau \right) } \\ 
\ \\ 
=1-\lim_{\tau \rightarrow \infty }\frac{1}{\mu H\left( \tau \right) }=1-%
\frac{1}{\mu H\left( \infty \right) }.%
\end{array}%
\right.  \label{A44}
\end{equation}%
In turn, Eq. (\ref{A44}) yields Eq. (\ref{43}).

Differentiating the middle part of Eq. (\ref{32}) yields%
\begin{equation}
\frac{M^{\prime }\left( \tau \right) }{\mu }=\frac{f_{res}\left( \tau
\right) F\left( \tau \right) -F_{res}\left( \tau \right) f\left( \tau
\right) }{F\left( \tau \right) ^{2}}.  \label{A45}
\end{equation}%
Eqs. (\ref{A45}) and (\ref{A40}) imply that if $\tau _{c}$ is a critical
point, $M^{\prime }\left( \tau _{c}\right) =0$, then%
\begin{equation}
F_{res}\left( \tau _{c}\right) =F\left( \tau _{c}\right) \frac{f_{res}\left(
\tau _{c}\right) }{f\left( \tau _{c}\right) }=\frac{F\left( \tau _{c}\right) 
}{\mu H\left( \tau _{c}\right) }.  \label{A46}
\end{equation}%
Substituting Eq. (\ref{A46}) into the middle part of Eq. (\ref{32}) yields 
\begin{equation}
\frac{M\left( \tau _{c}\right) }{\mu }=\frac{1}{F\left( \tau _{c}\right) }%
\left[ \frac{F\left( \tau _{c}\right) }{\mu H\left( \tau _{c}\right) }\right]
=\frac{1}{\mu H\left( \tau _{c}\right) }.  \label{A47}
\end{equation}%
In turn, Eq. (\ref{A47}) yields Eq. (\ref{44}).

\subsection{Derivation of Eqs. (\protect\ref{53})-(\protect\ref{54})}

Using Eq. (\ref{A40}) we have 
\begin{equation}
\left. 
\begin{array}{l}
F_{res}\left( \tau \right) -F\left( \tau \right) =\int_{0}^{\tau
}f_{res}\left( t\right) dt-\int_{0}^{\tau }f\left( t\right) dt \\ 
\ \\ 
=\int_{0}^{\tau }\left[ f_{res}\left( t\right) -f\left( t\right) \right]
dt=\int_{0}^{\tau }\left[ \frac{f_{res}\left( t\right) }{f\left( t\right) }-1%
\right] f\left( t\right) dt \\ 
\ \\ 
=\int_{0}^{\tau }\left[ \frac{1}{\mu H\left( t\right) }-1\right] f\left(
t\right) dt=\frac{1}{\mu }\int_{0}^{\tau }\left[ \frac{1}{H\left( t\right) }%
-\mu \right] f\left( t\right) dt.%
\end{array}%
\right.  \label{A52}
\end{equation}%
Substituting Eq. (\ref{A52}) into the middle part of Eq. (\ref{A80}) yields
Eq. (\ref{53}).

Using Eq. (\ref{A40}) we have%
\begin{equation}
\left. 
\begin{array}{l}
\bar{F}\left( \tau \right) -\bar{F}_{res}\left( \tau \right) =\int_{\tau
}^{\infty }f\left( t\right) dt-\int_{\tau }^{\infty }f_{res}\left( t\right)
dt \\ 
\ \\ 
=\int_{\tau }^{\infty }\left[ f\left( t\right) -f_{res}\left( t\right) %
\right] dt=\int_{\tau }^{\infty }\left[ 1-\frac{f_{res}\left( t\right) }{%
f\left( t\right) }\right] f\left( t\right) dt \\ 
\ \\ 
=\int_{\tau }^{\infty }\left[ 1-\frac{1}{\mu H\left( t\right) }\right]
f\left( t\right) dt=\frac{1}{\mu }\int_{\tau }^{\infty }\left[ \mu -\frac{1}{%
H\left( t\right) }\right] f\left( t\right) dt.%
\end{array}%
\right.  \label{A53}
\end{equation}%
Substituting Eq. (\ref{A53}) into the right part of Eq. (\ref{A80}) yields
Eq. (\ref{54}).

\subsection{Derivation of critical-timer results}

Taking logarithm on the left and middle parts of Eq. (\ref{32}), we
introduce the function%
\begin{equation}
L\left( \tau \right) =\ln \left[ \frac{M\left( \tau \right) }{\mu }\right]
=\ln \left[ F_{res}\left( \tau \right) \right] -\ln \left[ F\left( \tau
\right) \right] .  \label{A62}
\end{equation}%
($\tau >0$). Evidently, the local minima and the local maxima (if such
exist) of the functions $M\left( \tau \right) $ and $L\left( \tau \right) $
occur at the very same points.

Differentiating Eq. (\ref{A62}) with respect to the timer $\tau $, using the
backward hazard function of the input $T$ (Eq. (\ref{61})), and using the
backward hazard function of the input's residual lifetime $T_{res}$, we
obtain that:%
\begin{equation}
L^{\prime }\left( \tau \right) =\frac{f_{res}\left( \tau \right) }{%
F_{res}\left( \tau \right) }-\frac{f\left( \tau \right) }{F\left( \tau
\right) }=B_{res}\left( \tau \right) -B\left( \tau \right)  \label{A63}
\end{equation}%
($\tau >0$). Eq. (\ref{A63}) implies that%
\begin{equation}
L^{\prime }\left( \tau _{c}\right) =0 \\ \Leftrightarrow \\
B\left( \tau _{c}\right) =B_{res}\left( \tau _{c}\right) .
\label{A64}
\end{equation}%
In turn, Eq. (\ref{A64}) implies Eq. (\ref{62}).

Note that%
\begin{equation}
\left. 
\begin{array}{l}
\left[ \frac{f\left( \tau \right) }{F\left( \tau \right) }\right] ^{\prime }=%
\frac{f^{\prime }\left( \tau \right) F\left( \tau \right) -f\left( \tau
\right) f\left( \tau \right) }{F\left( \tau \right) ^{2}} \\ 
\ \\ 
=\frac{f^{\prime }\left( \tau \right) }{F\left( \tau \right) }-\left[ \frac{%
f\left( \tau \right) }{F\left( \tau \right) }\right] ^{2} \\ 
\ \\ 
=\frac{f^{\prime }\left( \tau \right) }{f\left( \tau \right) }\frac{f\left(
\tau \right) }{F\left( \tau \right) }-\left[ \frac{f\left( \tau \right) }{%
F\left( \tau \right) }\right] ^{2}%
\end{array}%
\right.  \label{A65}
\end{equation}%
($\tau >0$). Using the backward hazard function (Eq. (\ref{61})) and the
Gibbs gradient function (Eq. (\ref{63})) of the input $T$, Eq. (\ref{A65})
yields%
\begin{equation}
B^{\prime }\left( \tau \right) =-G\left( \tau \right) B\left( \tau \right)
-B\left( \tau \right) ^{2}  \label{A66}
\end{equation}%
($\tau >0$). Identically to Eq. (\ref{A66}), we have%
\begin{equation}
B_{res}^{\prime }\left( \tau \right) =-G_{res}\left( \tau \right)
B_{res}\left( \tau \right) -B_{res}\left( \tau \right) ^{2}  \label{A67}
\end{equation}%
($\tau >0$).

Differentiating Eq. (\ref{A63}) with respect to the timer $\tau $ yields 
\begin{equation}
L^{\prime \prime }\left( \tau \right) =B_{res}^{\prime }\left( \tau \right)
-B^{\prime }\left( \tau \right)  \label{A61}
\end{equation}%
($\tau >0$). Substituting Eqs. (\ref{A66}) and (\ref{A67}) into Eq. (\ref%
{A61}) further yields%
\begin{equation}
\left. 
\begin{array}{l}
L^{\prime \prime }\left( \tau \right) =\left[ -G_{res}\left( \tau \right)
B_{res}\left( \tau \right) -B_{res}\left( \tau \right) ^{2}\right] -\left[
-G\left( \tau \right) B\left( \tau \right) -B\left( \tau \right) ^{2}\right]
\\ 
\ \\ 
=\left[ G\left( \tau \right) B\left( \tau \right) -G_{res}\left( \tau
\right) B_{res}\left( \tau \right) \right] +\left[ B\left( \tau \right)
^{2}-B_{res}\left( \tau \right) ^{2}\right]%
\end{array}%
\right.  \label{A68}
\end{equation}%
($\tau >0$). In particular, for timers $\tau _{c}$ -- which satisfy Eq. (\ref%
{A64}) -- Eq. (\ref{A68}) implies that%
\begin{equation}
L^{\prime \prime }\left( \tau _{c}\right) =B\left( \tau _{c}\right) \left[
G\left( \tau _{c}\right) -G_{res}\left( \tau _{c}\right) \right] .
\label{A69}
\end{equation}%
Hence, for a critical timers $\tau _{c}$, Eq. (\ref{A69}) yields the two
following conclusions. (I) If $G\left( \tau _{c}\right) <G_{res}\left( \tau
_{c}\right) $ then a local maximum of the function $L\left( \tau \right) $
is attained at the timer $\tau _{c}$. (II) If $G\left( \tau _{c}\right)
>G_{res}\left( \tau _{c}\right) $ then a local minimum of the function $%
L\left( \tau \right) $ is attained at the timer $\tau _{c}$. In turn, these
results imply the local-maximum and the local-minimum results of section \ref%
{6}.

\end{document}